# Predicting Retrosynthetic Reaction using Self-Corrected Transformer Neural Networks

Shuangjia Zheng[1,2#], Jiahua Rao[2#], Zhongyue Zhang[2], Jun Xu*[1,3], and Yuedong Yang*[2,4]

[1]Research Center for Drug Discovery, School of Pharmaceutical Sciences, Sun Yat-sen University, 132 East Circle at University City, Guangzhou 510006, China

[2]School of Data and Computer Science, Sun Yat-sen University, Guangzhou 510006, China

[3]School of Computer Science & Technology, Wuyi University, 99 Yingbin Road, Jiangmen 529020, China

[4]Key Laboratory of Machine Intelligence and Advanced Computing, Sun Yat-sen University, Ministry of Education, Guangzhou 510000, China

## Abstract

Synthesis planning is the process of recursively decomposing target molecules into available precursors. Computer-aided retrosynthesis can potentially assist chemists in designing synthetic routes, but at present it is cumbersome and provides results of dissatisfactory quality. In this study, we develop a template-free self-corrected retrosynthesis predictor (SCROP) to perform a retrosynthesis prediction task trained by using the Transformer neural network architecture. In the method, the retrosynthesis planning is converted as a machine translation problem between molecular linear notations of reactants and the products. Coupled with a neural network-based syntax corrector, our method achieves an accuracy of 59.0% on a standard benchmark dataset, which increases >21% over other deep learning methods, and >6% over template-based methods. More importantly, our method shows an accuracy 1.7 times higher than other state-of-the-art methods for compounds not appearing in the training set.

# Introduction

Organic synthesis is one of the fundamental pillars of modern chemical society, as it provides a wide range of compounds from medicines to materials. The synthetic route to a desired organic compound is widely constructed by recursively decomposing it into a set of available reaction building blocks.[1] This analysis mode was formalized as retrosynthesis by E. J. Corey[2-3] who ultimately got the Nobel Prize in 1990.[4]

Planning synthesis requires chemists to predict functional groups reacting with a given reactant and their reacting poses. Since molecules may have many possible ways to decompose, the retrosynthetic analysis of a target compound usually leads to a large number of possible synthetic routes. It is challenging to select an appropriate synthesis route because the differences between routes are subtle that often depend on the global structures. Therefore, it remains challenging even for the best chemists to plan a retrosynthetic route for a complex molecule.[5-6]

To this end, many *in silico* methods have been developed to assist in designing synthetic routes for novel molecules, among which most are dependent on hand-coded reaction templates.[7-13] Based on these templates, synthesis routes were then built according to generalized reaction rules. Therefore, the accuracy of these methods depends on the availability of both templates and reaction rules. The rule-based systems required an extensive and up-to-date rule-base to cover the majority of known synthetic approaches, but such system could cover only a small fraction of chemist's knowledge base due to the constant increase in the number of new reactions.[14] Additionally, a simple template is generally not enough to reliably predict reactions, because it only identifies reaction centers and their neighboring atoms without considering the global information of the target molecule.

Recently, the so-called focused template-based methods were designed through automatically extracting templates from the reaction databases and applying the rules to selected relevant templates. It is critical to select appropriate templates. Segler and Waller employed neural networks to score template relevance based on molecular fingerprints.[15-16] Later, they showed that the Monte Carlo tree combined with deep neural networks could prioritize templates and pre-select the most promising retrosynthetic steps.[5] Coley and co-workers also demonstrated an approach for automated retrosynthesis based on analogy to known reactions.[17] Albeit sturdy, these methods rely heavily on pre-defined atom mapping to map atoms in the reactants to atoms in the product, which is still a nontrivial problem.[18-19] Besides, commonly used tools to identify the atom-mapping are themselves based on databases of expert rules and templates, which seems to get stuck in an infinite loop.[20] At any rate, template-based models have the limitation that they cannot infer reactions outside the chemical space covered by the template's libraries, and thus are restrained from discovering novel chemistry.[16]

To overcome the problem, template-free alternatives have emerged over recent years. The key idea is to use a text representation of the reactants and products (like SMILES), and thus convert retrosynthesis prediction as machine translation from one language (reactants) to the other language (products). Nam and Kim first described a neural sequence-to-sequence (seq2seq) model for the forward reaction prediction task.[21] Later, Liu and co-workers reported a similar seq2seq model that had comparable performance to a rule-based expert system.[22] This seq2seq model can be trained in a fully end-to-end fashion that does not require atom-mapped reaction examples for training. However, it doesn't show significant improvement in accuracy compared to the rule-based system and produces a great number of chemically invalid outputs. Though the invalid outputs can be easily identified according to chemical rules, there are no effective ways to correct the mistakes.

More recently, Transformer architecture shows advantages in machine translation.[23] It removes the traditional recurrent units and is based entirely on self-attention mechanism, allowing extracting both the local and global features without regard to their distance between the input and output sequences. Schwaller and co-workers employed this model to predict the products of chemical reactions and reached state-of-the-art results.[24] This indicates that Transformer architecture has the potential to perform the retrosynthetic reaction prediction task, though it is more challenging compared to the forward one.

In this study, we propose a novel template-free self-corrected retrosynthesis predictor (SCROP) built on the multi-head attention Transformer architecture. Our model achieves 59.0% top-1 accuracy on a standard benchmark dataset, which outperforms all the state-of-the-art template-free and template-based algorithms. At the same time, the rates of invalid candidate precursors could reduce from 12.1% to 0.7% by coupling with a novel neural network-based syntax checker. When excluding similar reactants from the training set, our method achieves an accuracy of 47.6% that is 1.7 times higher than other methods. More importantly, this model requires no handcrafted templates and atom mappings, and can accurately predict subtle chemical disconnections.

## Overview

**Dataset.** We have used the same reaction data as first parsed by Lowe.[25] The dataset was derived from USPTO granted patents that includes 50, 000 reactions that was later classified into 10 reaction classes by Schneider et al,[26] namely USPTO-50K. Figure 1 shows the distribution of each reaction class within the USPTO-50K. This dataset was also employed by Liu et al. and Coley et al. for the same task.[17, 22] We follow their random split strategy, with 40K, 5K, and 5K for training, validating, and testing.

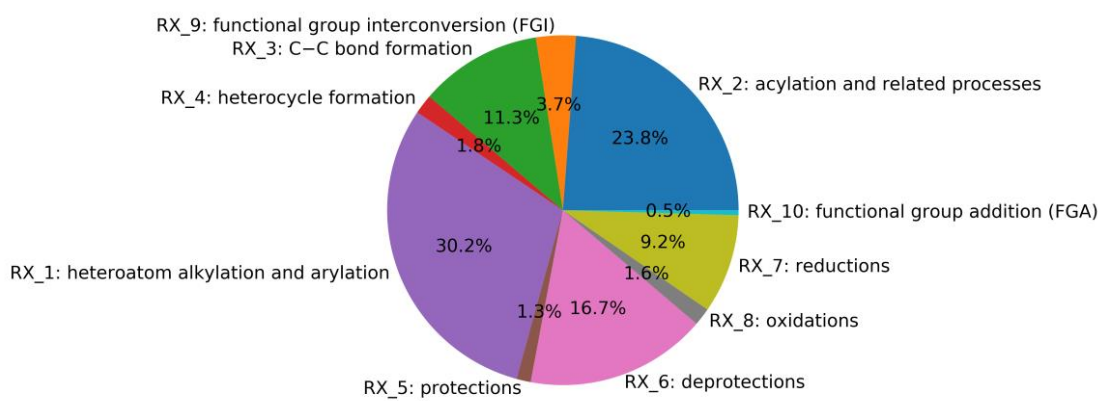

**Figure 1.** Distribution of Reaction Classes within the USPTO-50K.

The random split of the dataset that may separate similar reactants into the training and test set made the results over-estimated. In order to accurately estimate the performance for unseen compounds (compounds not covered by the chemical space of training data), we re-constructed a more challenging dataset using cluster split strategy as also used in the previous study,[27] where the compounds belonging to the same cluster were put to the same subset during the splitting. Here, we used 2-D similarity fingerprints to measure the topological similarities between target products, and employed the typically used Bemis−Murcko atomic frameworks[28] to cluster the products with a similarity threshold of 0.6. The same proportion (80%/10%/10%) was used to split the reactions into training, validation, and test set, namely USPTO-50K-cluster. The splitting makes the retrosynthesis prediction task significantly harder, as the model has to determine the reaction center for a target molecule outside its training (with a low similarity).

**Problem Definition.** Given an input of a target molecule and its specified reaction type, our task is to predict likely reactants that can react in the specified reaction type to form the target product. In this study, a reaction is described by a variable-length string containing one pair of SMILES notations representing the reactant and target compound. Following the process of Liu et al,[22] each reaction is split into a source sequence and target sequence for model training. The source sequence is the product of the reaction with a reaction type token prepended to the sequence, and the target sequence is the reactant set. For example, a protection reaction can be described as "NCc1ccoc1.S=(Cl)Cl>>[RX_5]S=C=NCc1ccoc1", where "[RX_5]S=C=NCc1ccoc1" is the source sequence, and "NCc1ccoc1.S=(Cl)Cl" is the target sequence. The "[RX_5]" token denotes the reaction class 5 (protections).

These sequences are then encoded into one-hot matrices with a token vocabulary (in our case, it has a totally 50 unique tokens retrieved from the dataset). In the one-hot encoding approach, each sequence is represented by a set of token vectors. All token vectors have the same number of components. Each component in a vector is set to zero except the one at the token's index position. To make the training procedure more stable and efficient, the input one-hot matrices are compressed to information-enriched word embedding vectors following the previous work.[29-30] As a result, each input sequence is finally represented in a molecular embedding:

$$m = (t_1, t_2, \ldots t_n) \tag{1}$$

where $t_i$ is a vector standing for a $d$ dimensional token embedding for the $i$-th token in a molecule consisted of $n$ tokens.

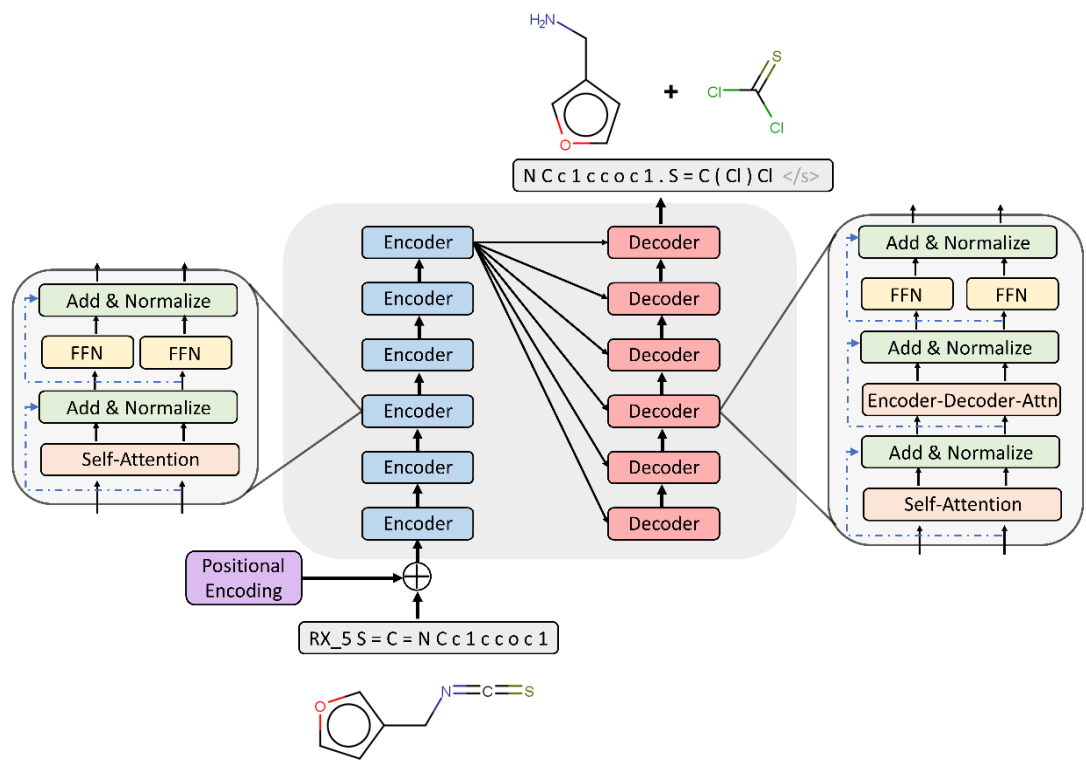

**Figure 2.** Overview of the architecture and training procedure of the Transformer-based retrosynthetic reaction predictor. 'RX_5' token indicates that the target should be decomposed through a protection reaction.

**Model**

Our model predicts the synthetic route for a target molecule in a two-step manner: (1) applying a fully-trained retrosynthetic reaction predictor to infer a set of raw candidate reactants, and (2) fixing their syntax errors to make more reasonable predictions using molecular syntax corrector.

**Retrosynthesis predictor.** In the first stage, we adapted the Transformer architecture to map the sequence of products to the sequence of the reactants.[23] As shown in Figure 2, the architecture of the Transformer system follows the so-called encoder-decoder paradigm, trained in an end-to-end fashion. The encoder layers attend the source molecular embedding $m_s = (t_1, \ldots, t_n)$ and iteratively transform it into a latent representation $l = (l_1, \ldots, l_n)$. After finishing the encoder phasing, each step in the decoding phase outputs a token based on the latent information $l$ until the ending token '</s>' is reached indicating that the transformer decoder has completed its output. The predicted output $y_p = (y_1, \ldots, y_m)$ is then used to compare with target reactants sequence $m_r = (t_1, \ldots, t_k)$. The training goal is to minimize the gap between $y_p$ and $m_r$ so that the model can finally infer accurate reactions.

Several identical layers are stacked for the encoding phase. Each layer comprises a combination of a multi-head self-attention sub-layer and a positional feedforward network (FFN) sub-layer. A residual connection and layer normalization were employed to integrate the two sub-layers.[31]

Different from the encoder, the decoder is composed of two types of attention multi-head attention layers: i) a decoder self-attention and ii) an encoder-decoder attention. The decoder self-attention focuses on the previous predictions of reactants made step by step, masked by one position. The encoder-decoder attention builds the connection between the final encoder representation and the decoder representation. It integrates the information of the source molecular embeddings with the reactants strings that have been predicted so far, which helps the decoder focus on appropriate places in the input sequence.

A multi-head attention unit itself comprised several scaled-dot attention layers performing the attention mechanism in parallel, which are then concatenated and projected to the final values. The scaled-dot attention layers take three matrices: the queries $Q$, the keys $K$, and the values $V$. The query, key, and value matrices are created by multiplying the input molecular embedding $M$ by three weight matrices that were also trained during the training process. We then compute the attention weight for each token within a SMILES string as follows:

$$Attention(Q, K, V) = softmax\left(\frac{QK^T}{\sqrt{d_k}}\right)V \qquad (2)$$

The dot-product of the keys and the queries computes how closely the keys are correlated with the queries. If the query and key are aligned well, their dot-product will be large. Each key has a corresponding value vector, which is multiplied with the output of the *softmax*. $d_k$ denotes a scaling factor depending on the weight matrices size. By this procedure, the encoder extracts pivotal features from the source sequence, which are then queried by the decoder depending on its preceding outputs. Thus, the model can learn the global level information from the input molecular embeddings and build a semantic connection between encoder and decoder.

As the recurrent unit is removed from the transformer architecture, the model lacks a way to account for the order of words in the input SMILES strings. To address this, we used the position encoding as proposed in the previous study,[23] which adopted the sine and cosine functions to identify the position of different tokens in the sequence:

$$PE_{(pos,2i)} = sin(\frac{pos}{timescale^{2i/d_{emb}}}), \quad PE_{(pos,2i+1)} = cos(\frac{pos}{timescale^{2i/d_{emb}}}) \quad (3)$$

where *pos* is the position, $i$ denotes the dimensional index of position encodings. The outputs of positional encodings have the same dimension $d_{emb}$ as the token embeddings. The timescale is set to 10000 to form a geometric progression from $2\pi$ to $10000 \cdot 2\pi$.

For a particular source sequence, the training objective is to minimize the cross-entropy loss function:

$$\mathcal{L}(y, m) = -\sum_{i=1}^{K} y_i \log(m_i) \quad (4)$$

where $y$ denotes the predicted sequence and $m$ is the target molecular sequence.

Our best-performance model was trained for 12 hours on one GPU (Nvidia 1080TI) on the training set, saving one checkpoint every 20, 000 steps and averaging the last ten checkpoints. More detailed hyperparameter settings of the model are shown in Table S1. The hyperparameters were chosen according to performances on the validating set. A beam search procedure[32] is then used to infer multiple reactant candidates on the test set. We used the best-performance model to infer the reactant candidates with a beam width of 10. Therefore, the top ten candidate sequences ranked by overall probability are retained.

**Molecular syntax corrector.** It is important to note that syntactically plausible molecular strings are not guaranteed to be semantically valid. For instance, 'c1ccoc' cannot be deduced to a valid structure because it misses the token '1' representing the closing of the heterocycle. Previous works entirely relied on the raw outputs obtained from the default beam search.[22, 33] This procedure enumerates the top $N$ predictions based on the joint probability of generated tokens without consideration of chemical feasibility. Thus, in the second stage, we build a Transformer-based syntax corrector to automatically correct the syntax of unreasonable SMILES strings for improving the model performance. The design of the neural network is motivated by the grammatical error correction tool widely utilized in natural language processing tasks.[34] Figure 3 illustrates the procedure of our syntax correction system. The syntax corrector takes the unreasonable predictions generated from the retrosynthetic reaction predictor and fixes their syntax errors to increase the quality of predictions.

The system does this by taking ground truth reactants and invalid reactants generated from the retrosynthesis predictor to produce input-output pairs (where the output is the ground truth reactants), which are then used to train a sequence-to-sequence transformer model. Concretely, we first use a fully trained model to generate the top ten candidate precursors given a set of target compounds in the training set. Second, we filter the candidate reactant sets by removing the ground truth reactant sets corresponding to the target molecules. Third, we construct a training library that consists of a set of input-output pairs, where the inputs are predicted invalid reactants, and the outputs are the ground truth reactants. Given such a syntax correction dataset with input-output pairs, we can train a new sequence-to-sequence model using the Transformer architecture introduced above and hook it up to the retrosynthetic reaction predictor. Once trained, we can input unsatisfactory SMILES strings generated from the reaction predictor and fix their syntax errors to make more reasonable predictions. Note that we only retained the top-1 candidate produced by syntax corrector and replace the original one.

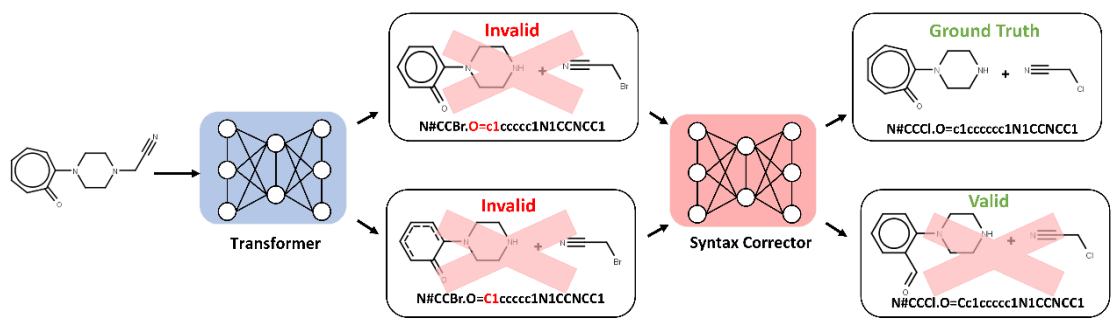

**Figure 3.** Example SMILES syntax correction for two invalid predictions generated by transformer-based retrosynthesis predictor. The syntax corrector fixes the syntax errors and produces the ground truth reactants.

After correction, we canonicalize all predicted sequences by reordering the tokens with fixed rules and compared the predicted candidates with the ground truth reactants. All scripts were written in Python (version 3.6), and RDKit was used for data preprocessing and molecule canonicalization.[35] The Transformer model was implemented using OpenNMT.[36]

## Experiments and results

We evaluated the self-corrected retrosynthesis predictor (SCROP) on USPTO-50K data sets with two split methods (random and clustering) and compared the performance with other state-of-the-art results (we have repeated their results as reported in their original papers). Table 1 shows the quantitative performance of the models on the USPTO-50K dataset when the reaction type is known and unknown, respectively. When inferring reaction within a specific reaction class, the SCROP outperforms all baselines in the top-1 recommendation, achieving an exactly matching accuracy of 59.0%. The percentages of correctly predicted reactants by top-3, top-5, and top-10 are 74.8%, 78.1%, and 81.1%, respectively. These results are much better than the results reported in the previous seq2seq model (37.4% of top-1 and 61.7% of top-10).

Note that the template-based model of Coley et al. (similarity) predicted 100 candidates and picked the top-10 as a result. As the SCROP does not rank candidates but was trained on accurately predicting the top-1 outcome and only predicted ten candidates, it is not surprising that the similarity-based method has higher top-5 and top-10 accuracies. Even so, the SCROP improves the similarity-based method by a margin of 6.1% in top-1 accuracy.

By comparing with the results of the original retrosynthesis predictor (SCROP-noSC), we find that the syntax corrector leads to an increase from top-1 to top-10 accuracies, ranging from 0.2% to 1.0%. The growth of top-1 and top-3 accuracy are not apparent because only a small part of invalid predictions was generated in these two stages.

Without prior knowledge of the reaction class (removing the reaction type tokens in the training procedure), the SCROP improves upon the similarity-based method by a margin of 6.4%, 5.7% in top-1, top-3 accuracies, and achieves a comparable result in top-5 and top-10 suggestions.

**Table 1.** Comparison of Top-N accuracies between the baselines and SCROP on USPTO-50K.

| Data | model | top-n accuracy (%), n = | | | |
|---|---|---|---|---|---|
| | | 1 | 3 | 5 | 10 |
| With reaction class | Liu-baseline | 35.4 | 52.3 | 59.1 | 65.1 |
| | Liu-seq2seq | 37.4 | 52.4 | 57.0 | 61.7 |
| | similarity | 52.9 | 73.8 | **81.2** | **88.1** |
| | SCROP-noSC | 58.8 | 74.4 | 77.5 | 80.1 |
| | SCROP | **59.0** | **74.8** | 78.1 | 81.1 |
| Without reaction class | similarity | 37.3 | 54.7 | 63.3 | **74.1** |
| | SCROP-noSC | 43.3 | 59.1 | 64.0 | 67.0 |
| | SCROP | **43.7** | **60.0** | **65.2** | 68.7 |

**Generalization Estimation.** To compare the generalization ability of these approaches, we further evaluated the models on USPTO-50K-cluster. We retrained both the seq2seq and similarity-based methods on the USPTO-50K-cluster with the same parameter settings as in the original papers, except the generated candidates in the similarity-based method was set to 10 rather than 100 for a fair comparison. Table 2 shows the models' performance aggregated across all classes on the USPTO-50K-cluster dataset. By this table, we observe that the SCROP achieved an order-of-magnitude improvement over baselines within or without reaction class from top-1 to top-10 predictions. Besides, we find that our model performs more stable compared to the template-based method. The top-1 accuracy of our model decreases only by 7.5% compared to 16.6% of the similarity method when planning retrosynthesis for clustered dataset without the knowledge of reaction class. This result demonstrates that our template-free method has better generalization ability than the template-based one when the training dataset has no similar compound to the target compounds.

**Table 2.** Comparison of top-N accuracies between the baselines and SCROP on USPTO-50K-cluster.

| Data | model | top-n accuracy (%), n = | | | |
|---|---|---|---|---|---|
| | | 1 | 3 | 5 | 10 |
| With reaction class | seq2seq | 25.5 | 38.7 | 43.6 | 49.0 |
| | similarity | 36.7 | 58.0 | 61.4 | 67.2 |
| | SCROP | **47.6** | **63.9** | **68.1** | **71.1** |
| Without reaction class | seq2seq | 16.5 | 28.8 | 34.0 | 40.6 |
| | similarity | 20.7 | 36.3 | 43.2 | 46.9 |
| | SCROP | **36.2** | **52.0** | **57.1** | **60.9** |

The detailed top-10 results on USPTO-50K-cluster dataset for the baseline models and our model broken down by the reaction classes are shown in Table 3. The SCPOR performs significantly better than the seq2seq model in reaction class 4 (heterocycle formation). The common feature of this reaction type is the formation of cyclic structures, resulting in a significant difference between the reactant SMILES string and the target product SMILES string. This result shows that the SCPOR is able to induce better syntactic relationships and capture global chemical information from the reaction data. The SCPOR also outperforms the seq2seq and similarity models by a large margin in class 1 (heteroatom alkylation and arylation). The key feature of this reaction class is that the reactions are possibly happened with many different functional groups within the target molecules. It is hard to identify the accurate reaction site when the structure of target molecules is not similar to the one in the knowledge base. As a result, the SCPOR shows better generalization ability in this reaction type and infers reactants correctly.

**Table 3.** Model top-10 accuracy within each class on USPTO-50K-cluster when the reaction type is known.

| model | top-10 accuracy (%), reaction class= | | | | | | | | | |
|---|---|---|---|---|---|---|---|---|---|---|
| | 1 | 2 | 3 | 4 | 5 | 6 | 7 | 8 | 9 | 10 |
| seq2seq | 43.9 | 58.7 | 29.8 | 11.9 | 63.5 | 54.4 | 66.0 | 54.4 | 43.3 | 45.8 |
| similarity | 56.4 | 77.0 | 50.8 | 52.4 | **86.2** | **77.5** | 73.6 | **86.2** | 58.8 | **85.3** |
| SCROP | **71.8** | **78.9** | **57.2** | **56.2** | 85.1 | 68.5 | **79.9** | 69.6 | **68.0** | 68.8 |

**Evaluation of syntax corrector.** In Liu's work,[7] incorrect predictions were summarized as three types: grammatically invalid outputs, grammatically valid but chemically unreasonable, and chemically plausible but do not match to the ground truth. In the above comparison, we only used the syntax corrector to fix the grammatically invalid outputs since the identification of the other two types of errors requires the knowledge of ground truth reactants. As shown in Figure 4, for SCROP, only 0.7% of the top-1 and 2.3% of the top-10 predictions are grammatically invalid, which is significantly better than the retrosynthesis predictor (SCROP-noSC) and seq2seq model. Figure 5 shows an example of retrosynthetic predictions coupling with the syntax corrector. The model successfully proposes the ground truth by correcting the original rank 2 prediction, which is an invalid molecule string. Besides, the syntax corrector makes the rank 3 prediction more reasonable, even if it is not the correct answer. This result suggests that the syntax corrector can be used to optimize the invalid predictions. It provides a new strategy to optimize the synthesis route when chemists find that the reaction is going to a wrong way.

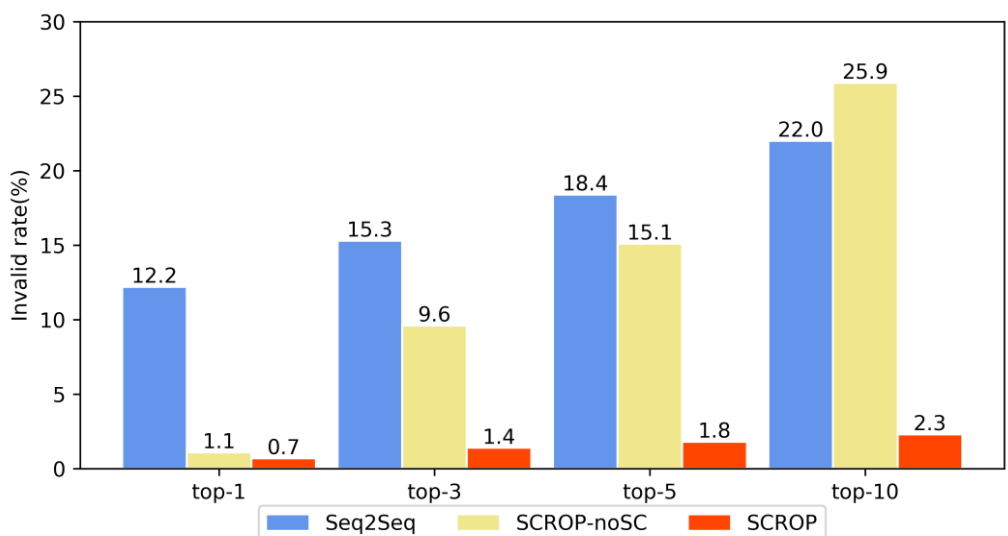

**Figure 4.** Comparison of invalid rates among seq2seq, retrosynthesis predictor (SCROP-noSC) and self-corrected retrosynthesis predictor (SCROP) for different beam sizes.

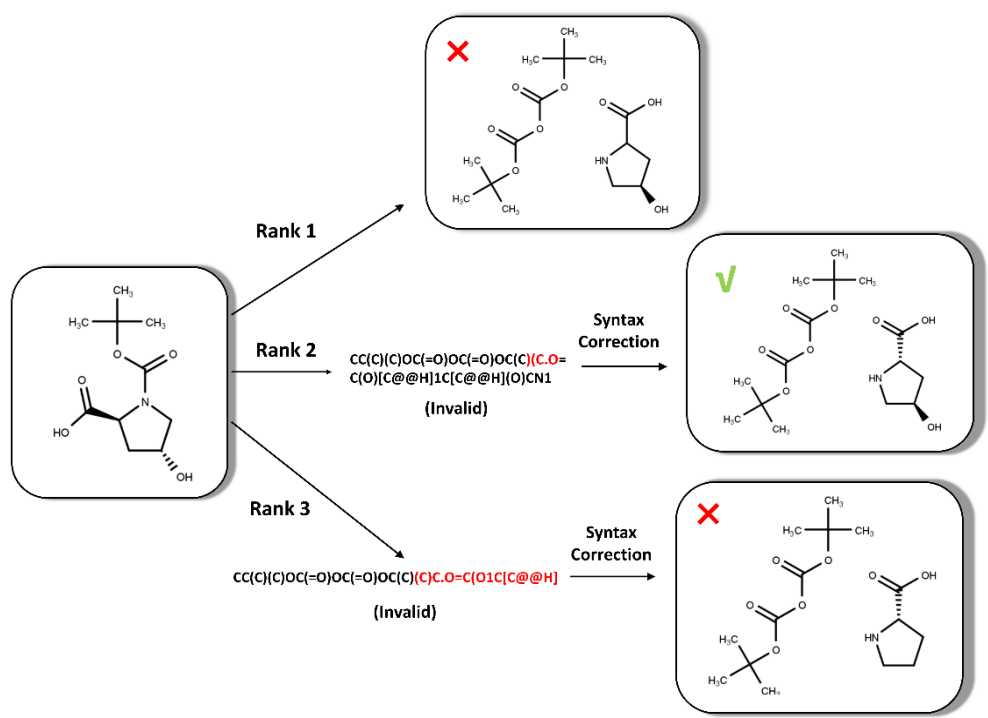

**Figure 5.** Example of retrosynthetic predictions coupling with the syntax corrector. The model successfully proposes the ground truth by correcting the original rank 2 prediction, which is an invalid molecule string. Red color denotes incorrect grammar.

**Attention analysis.** To investigate what the model has learned, we further analyzed the attention weights in our model. The attention weights provide clues on tokens in the input string that were considered to be more critical when a particular symbol in the output string was generated. Figure 6, for example, shows the top-1 candidate's attention maps of an acylation reaction extracted from the SCROP (accurately predicted) and seq2seq (wrong predicted) models. We observe that strong weights trend diagonally in SCROP's attention map, which constantly align SMILES substrings that are shared between the input and output. Besides, when inferring the unseen reaction sites 'OH' and '.', the model can simultaneously take several non-continuous tokens into account. This suggests that the SCROP tries to extract both the local and global information of the source sequence. In comparison, the attention map of the seq2seq model fails to align the substrings of the input-output pair and results in a wrong prediction.

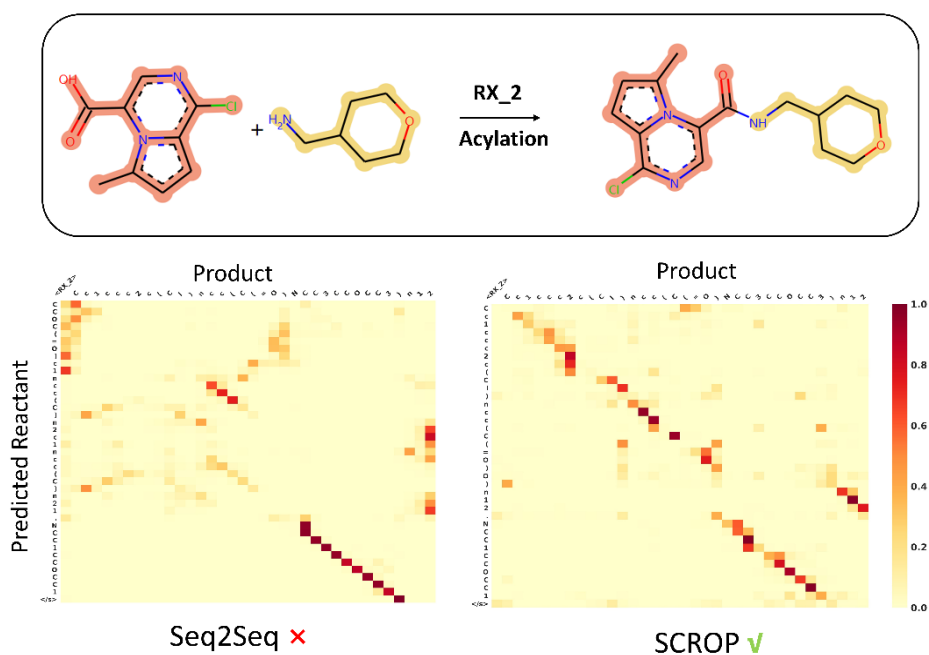

**Figure 6.** Top-1 candidate's attention maps of an acylation reaction extracted from the SCROP (accurately predicted) and seq2seq (wrong predicted) models.

**Limitation.** A distinct disadvantage is that the model scores candidates only taking account of the overall probability of the predicted strings. Practically, there are many additional considerations in synthetic route design, such as process complexity, expense, productivity, safety, and so on. This is because the public data sets do not contain additional technical information except for the reaction strings itself. We would explore the scoring system by incorporating additional considerations like similarity, reaction complexity, cost, and reaction yield to optimize the present evaluation metric.

Another limitation of the model is the multi-step reaction. A possible option would be to recursively decompose the target compound using the transformer prediction system until the commercially available building block is obtained. Monte Carlo tree search could be employed to score the outputs in each single reaction step.[5]

Besides, finding the best-performing set of hyperparameters for a deep neural network as well as the inference procedure is computationally expensive, as many hyperparameter settings take tens of hours to train using one GeForce GTX1080Ti graphics card. For this reason, we use a rough grid search to identify the final settings in training and inferring procedure. Better optimization techniques with sufficient equipment may result in a further increase in accuracy.

## Conclusions

In this work, we have proposed a novel template-free deep learning method for chemical retrosynthetic prediction. Instead of generating candidate precursors by reaction templates, we employ Transformer neural networks to generate potential candidates by learning the sequential representation of reactant-product pairs. Our model achieves 59.0% top-1 accuracy on a standard benchmark dataset, which outperforms all the state-of-the-art template-free and template-based algorithms. At the same time, the rates of invalid candidate precursors could reduce from 12.1% to 0.7% by coupling with a novel neural network-based syntax checker. When excluding similar reactants from the training set, our method achieves an accuracy of 47.6% that is 1.7 times higher than other methods. More importantly, our method is trained in an end-to-end fashion and is free of chemical rules, and the accuracy will improve automatically with the increase of the training samples.

**Acknowledgment**

The work was supported in part by the National Key R&D Program of China (2018YFC0910500), GD Frontier & Key Tech Innovation Program (2019B020228001), the National Natural Science Foundation of China (61772566, U1611261 and 81801132), and the program for Guangdong Introducing Innovative and Entrepreneurial Teams (2016ZT06D211).

**Author information**

*To whom correspondence should be addressed.

Yuedong Yang: yangyd25@mail.sysu.edu.cn

Jun Xu: junxu@biochemomes.com

**Competing interests**

The authors declare that they have no competing interests

**Authors' contributions**

SZ, JR, and YY contributed concept and implementation. SZ and JR co-designed experiments. SZ and JR were responsible for programming. All authors contributed to the interpretation of results. SZ and YY wrote the manuscript. All authors reviewed and approved the final manuscript.